\documentclass[%
 reprint,
 amsmath,amssymb,
 aps,
 prx,
superscriptaddress,
longbibliography,
]{revtex4-1}

\usepackage{graphicx}% Include figure files
\usepackage{dcolumn}% Align table columns on decimal point
\usepackage{bm}% bold math
\usepackage{natbib}
\usepackage{color}
\usepackage{tabularx}
\usepackage{float}
\usepackage{upgreek}
\usepackage{nicefrac}

\begin{document}

\title{Highly enriched $^{28}$Si reveals remarkable optical linewidths and fine structure for well-known damage centers}% Force line breaks with \\

\author{C. Chartrand}
\affiliation{Department of Physics, Simon Fraser University, Burnaby, British Columbia, Canada V5A 1S6}
\author{L. Bergeron}
\affiliation{Department of Physics, Simon Fraser University, Burnaby, British Columbia, Canada V5A 1S6}
\author{K. J. Morse}
\affiliation{Department of Physics, Simon Fraser University, Burnaby, British Columbia, Canada V5A 1S6}

\author{H. Riemann}
\affiliation{Leibniz-Institut f\"ur Kristallz\"uchtung, 12489 Berlin, Germany}
\author{N. V. Abrosimov}
\affiliation{Leibniz-Institut f\"ur Kristallz\"uchtung, 12489 Berlin, Germany}

\author{P. Becker}
\affiliation{Physikalisch-Technische Bundestanstalt Braunschweig, 38116 Braunschweig, Germany}

\author{H.-J. Pohl}
\affiliation{VITCON Projectconsult GmbH, 07743 Jena, Germany}

\author{S. Simmons}
\affiliation{Department of Physics, Simon Fraser University, Burnaby, British Columbia, Canada V5A 1S6}
\author{M. L. W. Thewalt}
\thanks{thewalt@sfu.ca}
\affiliation{Department of Physics, Simon Fraser University, Burnaby, British Columbia, Canada V5A 1S6}

\date{\today}% It is always \today, today,

\setlength{\skip\footins}{0.5cm}

\begin{abstract}

Luminescence and optical absorption due to radiation damage centers in silicon has been studied exhaustively for decades, but is receiving new interest for applications as emitters for integrated silicon photonic technologies. While a variety of other optical transitions have been found to be much sharper in enriched $^{28}$Si than in natural Si, due to the elimination of inhomogeneous isotopic broadening, this has not yet been investigated for radiation damage centers. We report results for the well-known G, W and C damage centers in highly enriched $^{28}$Si, with optical linewidth improvements in some cases of over two orders of magnitude, revealing previously hidden fine structure in the G center emission and absorption. These results have direct implications for the linewidths to be expected from single center emission, even in natural Si, and for models for the G center structure. The advantages of $^{28}$Si can be readily extended to the study of other radiation damage centers in Si.

\end{abstract}

\maketitle

%\tableofcontents

\section{\label{sec:intro}Introduction}

A multitude of radiation damage centers in Si, with highly reproducible optical emission and absorption lines, have been studied exhaustively using a wide variety of techniques over the past 50+ years \cite{Davies1989}. Some of these centers produce very bright luminescence, and have recently received renewed interest as possible electrically-pumped light emitters, and single photon sources, compatible with an integrated silicon photonic technology \cite{Shainline2007, Buckley2017, Beaufils2018}. Previous high resolution studies of these centers in natural Si have shown reproducible limiting linewidths of typically greater than 0.04 meV, and it has become widely assumed that these relatively narrow linewidths represent some kind of fundamental limit for these centers. A wide variety of other optical transitions in Si have been shown to become remarkably sharper in $^{28}$Si than in natural Si ($^{\mathrm{nat}}$Si) due to the elimination of the inhomogeneous broadening resulting from the mixture of the three stable Si isotopes ($^{28}$Si, $^{29}$Si, $^{30}$Si) present in $^{\mathrm{nat}}$Si \cite{Karaiskaj2001,Cardona2005,Steger2009_2,Steger2012, Morse2017}, but until now no studies of radiation damage centers have been undertaken in highly enriched $^{28}$Si. While 

\onecolumngrid

\begin{center}
\begin{figure}[b]
\includegraphics[width=\linewidth]{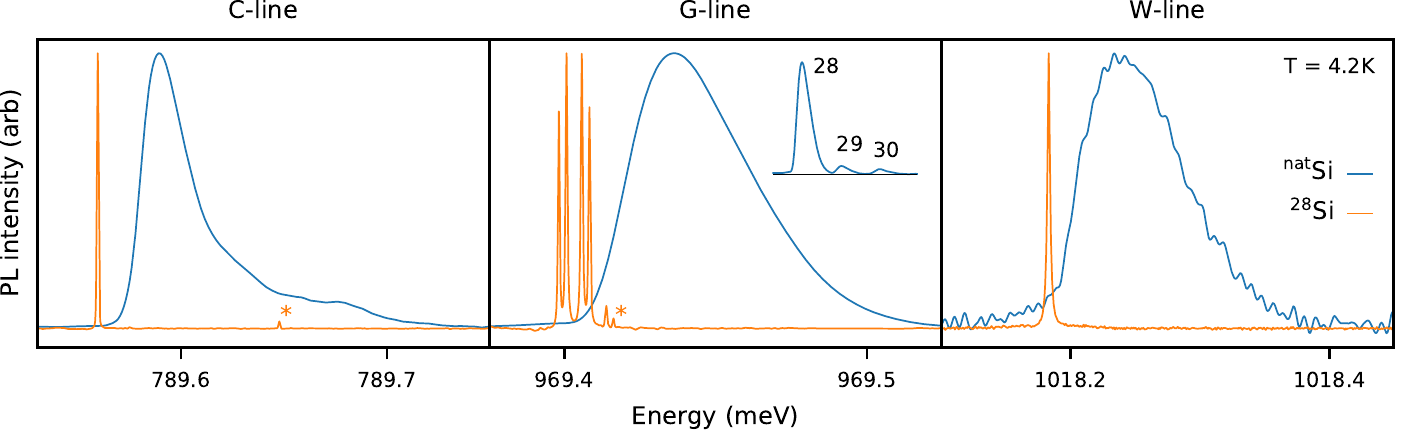}
\caption{Photoluminescence spectra of the C, G, and W no-phonon lines at \mbox{4.2 K} for samples irradiated with a \mbox{32 kGy} dose are compared for $^{\mathrm{nat}}$Si (blue) and $^{28}$Si (orange). The $^{\mathrm{nat}}$Si G-line spectrum is plotted in the inset with the energy range compressed by a factor of three to show the complete spectrum resulting from the three isotopes of Si occupying the interstitial site. The small features labeled with an * in the C and G $^{28}$Si spectra result from the replacement of one $^{12}$C in the center with a naturally occurring $^{13}$C. Note that the relative intensities of the $^{28}$Si G-line components are affected by reabsorption, as discussed in the text. All spectra were recorded with a resolution of \mbox{0.62 $\upmu$eV} ($^{28}$Si) and \mbox{6.2 $\upmu$eV} ($^{\mathrm{nat}}$Si), with the latter being sufficient to minimize instrumental contributions to the observed lineshape.}
    \label{fig:all}
\end{figure}
\end{center}
\twocolumngrid
\clearpage

\noindent there have been studies of radiation damage centers in $^{30}$Si, that material was not as highly enriched as the available $^{28}$Si, and in addition was grown by the Czochralski process, with the high oxygen concentration resulting in additional inhomogeneous broadening, so no linewidth improvements over $^{\mathrm{nat}}$Si were observed \cite{Hayama2004_1, Hayama2004_2}.\\

Here we report on optical emission and absorption studies on three well-known damage centers produced in very high purity Si enriched to 99.995$\%$ $^{28}$Si as part of the Avogadro project \cite{Becker2010}. All three emit in or near important optical communication bands: the C center (790 meV, or 1571 nm, thought to contain an interstitial C (C$_\mathrm{i}$) and an interstitial O (O$_\mathrm{i}$) \cite{Wagner1984,Davies1986, Davies1987_1,Trombetta1987,Kurner1989,Bohnert1993}), the G center (969 meV, or 1280 nm, thought to contain a substitutional C (C$_\mathrm{s}$), C$_\mathrm{i}$, and an interstitial Si (Si$_\mathrm{i}$) \cite{Davies1987_1,Canham1987,Song1990,Beaufils2018}), and the W center (1019 meV, or 1217 nm, thought to consist of three Si$_\mathrm{i}$ \cite{Davies1987_2,Davies2006, Surma2008}). All three produce strong photoluminescence (PL) in $^{28}$Si irradiated with 10 MeV electrons, and the G center also has relatively strong absorption transitions. All three no-phonon (NP) lines are much narrower in $^{28}$Si than ever observed in $^{\mathrm{nat}}$Si, by over two orders of magnitude for the C and G centers, and a factor of $\approx$50 for W. \\

In particular, the improved resolution of the G transition possible in $^{28}$Si reveals a quartet fine structure which is completely hidden in the inhomogeneously broadened linewidth seen for G centers in $^{\mathrm{nat}}$Si. The explanation of this quartet fine structure represents an important challenge for theoretical models of the G center \cite{Wang2014, Timerkaeva2017}. The linewidths of the G center components in $^{28}$Si are slightly over twice the homogeneous lifetime broadening limit. These ultranarrow ensemble linewidths, and the quartet fine structure of the G-line, have direct implications for the emission and absorption spectra of individual, isolated centers, even in $^{\mathrm{nat}}$Si, and for their possible use as single photon emitters. \\

It is interesting to note that in his 1989 review of luminescence centers in Si, Davies \cite{Davies1989} proposed that the limiting linewidths of centers such as G in the best $^\text{nat}$Si samples might result from inhomogeneous broadening due to the distribution of Si isotopes.  Our results finally demonstrate the prescience and generality of this proposal. \\

\section{\label{sec:exp}Experimental methods}

The $^{28}$Si samples used here were cut from the seed end of the Avogadro crystal \cite{Becker2010}, enriched to 99.995$\%$ $^{28}$Si and containing less than 5$\times$10$^{14}$ cm$^{-3}$ C and 1$\times$10$^{14}$ cm$^{-3}$ O, and irradiated at room temperature with either \mbox{32 kGy} or \mbox{320 kGy} doses of 10 MeV electrons. Control samples cut from high-purity undoped floating-zone-grown $^{\mathrm{nat}}$Si received the same irradiations. Individual samples were subsequently annealed for 30 min in air at temperatures between \mbox{100 $^\circ$C} and \mbox{250 $^\circ$C}, with \mbox{100 $^\circ$C} producing strong photoluminescence (PL) for the G center, and \mbox{250 $^\circ$C} for the C and W centers. All samples were etched in HF/HNO$_3$ prior to measurements in order to remove any surface damage. Samples were held in a strain free manner immersed in liquid He at either \mbox{4.2 K} or \mbox{1.4 K}, except for the temperature controlled experiments which were conducted in flowing He gas. Note that at the high spectral resolutions possible with $^{28}$Si, there are significant energy shifts between \mbox{4.2 K} (atmospheric pressure), and \mbox{1.4 K} (near zero pressure), resulting from the temperature and pressure dependence of the Si band gap energy \cite{Cardona2004}. The line energies given later are therefore all specified at \mbox{4.2 K} and atmospheric pressure.\\

All PL spectra, and some of the absorption spectra, were collected with a Bruker IFS 125 HR Fourier transform infrared (FTIR) spectrometer, using a CaF$_2$ beamsplitter, and a liquid nitrogen cooled Ge detector for PL, or a room temperature InGaAs detector and Tungsten-halogen source for absorption. For PL spectra, excitation was provided by either 200 mW of 532 nm radiation, or 300 mW of 1047 nm radiation, with the bulk excitation at 1047 nm providing a somewhat better signal-to-noise ratio (SNR). The resolution of the FTIR considerably exceeded the specification of 10$^6$, providing more than enough resolution to determine the FWHM of the W-line in $^{28}$Si, and contributing only a small broadening to the observed width of the G-line components, but was not quite sufficient to determine the FWHM of the C-line in our $^{28}$Si samples.\\

Higher resolution absorption spectra for the C and G centers in $^{28}$Si were obtained using single frequency temperature-controlled distributed feedback (DFB) semiconductor diode lasers, scanned by varying the diode current in steps as small as \mbox{1 $\upmu$A}, and calibrated using a high resolution wavemeter. For the G center, which had quite strong absorption in our samples, this could be done in a straightforward transmission geometry, using a room-temperature InGaAs detector. For the G center spectroscopy using the DFB laser, additional above-gap excitation at 1047 nm was also used to reverse the observed resonant bleaching of the G center absorption. A second DFB laser at the G center wavelength was also used to investigate selective bleaching of the G center absorption. The absorption due to the C center was too small to detect directly in our samples, but the absorption spectrum could be obtained by using photoluminescence excitation spectroscopy (PLE), in which the resonant absorption of the laser light could be detected by monitoring the broad TA phonon sideband of the C center emission using a 3/4 m focal length double spectrometer as a spectral filter, and a high-sensitivity InGaAs photon-counting detector. The spectral widths of the DFB lasers were specified to be less than 2 MHz, much narrower than the observed C or W linewidths.

\section{\label{sec:results}Results and discussion}

In the following three subsections we will discuss our results for the G, C and W centers, after first reviewing some of the relevant literature for each of these systems. A comparison of the PL spectra of the NP lines of these three centers in $^{28}$Si and $^{\mathrm{nat}}$Si is shown in \mbox{Fig. \ref{fig:all}}, demonstrating immediately the dominance of inhomogeneous Si isotope broadening in determining the linewidths observed in all $^{\mathrm{nat}}$Si spectra. Energies and linewidths of the observed NP components are given in \mbox{Table \ref{tab:PL}}. Also apparent in \mbox{Fig. \ref{fig:all}} is a peak energy downshift for all three centers in $^{28}$Si as compared to $^\mathrm{nat}$Si, ranging from 25\% to 50\% of the 114$\upmu$eV band gap shift between $^{28}$Si and $^\mathrm{nat}$Si \cite{Karaiskaj2001}. This energy shift with average host isotope mass will not be discussed further here as it has already been considered in detail by Hayama et al. \cite{Hayama2004_1,Hayama2004_2} in comparing the spectra of these defects in $^{\mathrm{nat}}$Si and $^{30}$Si.

\begin{table}[b]
  \centering
  \begin{tabularx}{\linewidth}{X X X X X}
  \hline \hline\noalign{\smallskip}
   & \multicolumn{2}{X}{Energy} & \multicolumn{2}{X}{FWHM} \\
  	   & \multicolumn{2}{X}{[meV]} & \multicolumn{2}{X}{[$\upmu$eV]} \\\noalign{\smallskip}\hline\noalign{\smallskip}
  	  	 & $^{28}\mathrm{Si}$ & $^{\mathrm{nat}}\mathrm{Si}$ & $^{28}\mathrm{Si}$ & $^{\mathrm{nat}}\mathrm{Si}$\\\noalign{\smallskip}\hline\noalign{\smallskip}
   C$_\mathrm{T}$ 	 & 786.9261 & 786.95  & &  \\    
    C$_0$ 			 & 789.5596 & 789.589 & 0.19(1) & 27(1)\\
    C$_0$ ($^{18}$O) & 789.5750 & 		  & &  \\
    C$_0$ ($^{13}$C) & 789.6477 & 789.676 & & \\\noalign{\smallskip}\hline\noalign{\smallskip}
    G$_1$               & 969.3982 & 969.436 & 0.23(2) & 46(1) \\ %4.24
    G$_2$               & 969.4007 &         & 0.23(2) &  \\ %5.47
    G$_3$               & 969.4057 &         & 0.23(2) &  \\ %5.52
    G$_4$               & 969.4083 &         & 0.25(2) &  \\  %4.45
    G$_{5a}$ ($^{13}$C) & 969.4135 &         & 0.3(1)   &  \\ %$<$3.1
    G$_{5b}$ ($^{13}$C) & 969.4139 &         & 0.3(1)   &  \\  %$<$3.1
    G$_{6a}$ ($^{13}$C) & 969.4160 &         & 0.3(1)  & \\ %$<$3.1
    G$_{6b}$ ($^{13}$C) & 969.4163 &         & 0.3(1)   &  \\\noalign{\smallskip}\hline\noalign{\smallskip}

    W & 1018.1829 & 1018.24 &  2.0(1) & 92(9) \\\hline \hline
  \end{tabularx}
    \caption{Energies and full width at half maximum (FWHM) of C, G and W no-phonon lines in $^{28}$Si from photoluminescence in the case of W, from photoluminescence excitation for C and from absorption for G. All energy positions are given for \mbox{4.2 K}. For the $^{\mathrm{nat}}$Si lines, which are asymmetrical, the peak energy is given. The absolute energy accuracy is \mbox{1 $\upmu$eV} for the $^{28}$Si results and \mbox{10 $\upmu$eV} for the $^{\mathrm{nat}}$Si results, but the relative accuracy, for example in determining the separation of the main G-line components in $^{28}$Si, is equal to \mbox{0.1 $\upmu$eV}.}

  \label{tab:PL}
\end{table}

\subsection{\label{subsec:G center}G center (969 meV, 1280nm)}

The G center, with a NP line at \mbox{$\sim$969 meV} \mbox{($\sim$1280 nm)} is one of the most studied radiation damage centers in Si due to its extraordinarily bright emission, as well as large oscillator strength as evidenced by strong optical absorption even for relatively low G center concentrations \cite{Davies1989}. Beaufils et al. \cite{Beaufils2018} provide an up-to-date and comprehensive review of the G center literature, together with new data on temperature dependences, and a precise measurement of the \mbox{5.9 ns} luminescence decay time. The G center is now widely agreed to consist of a substitutional-interstitial C pair (C$_\mathrm{i}$-C$_\mathrm{s}$) coupled to a Si$_\mathrm{i}$. Song et al. \cite{Song1990} have argued that the G center exists in two configurations, A and B, as well as positive, neutral and negative charge states, with the G luminescence assigned to the neutral B form. More recent modelling has proposed three \cite{Wang2014} and even four \cite{Timerkaeva2017} configurations for the G center, so its detailed understanding remains in a state of flux.\\

The formation pathway of the G center is similar to that of many other radiation damage centers and is relatively well understood \cite{Davies1989}. Irradiation, for example with high energy electrons, produces primarily isolated Si$_\mathrm{i}$ and Si vacancies, both of which are mobile at room temperature. These may annihilate with each other, complex with themselves, or interact with other impurities and defects present in the crystal. Of central importance to the formation of many of these centers is the replacement of C$_\mathrm{s}$ by Si$_\mathrm{i}$ to produce C$_\mathrm{i}$, which can be directly detected by a variety of probes \cite{Davies1989}. C$_\mathrm{i}$ is somewhat mobile at room temperature and becomes increasingly mobile at slightly elevated temperatures, allowing it to combine with, for example, C$_\mathrm{s}$, leading to formation of the G center, or O$_\mathrm{i}$, leading to formation of the C center. Other centers produced after higher temperature annealing may form from the dissociation products of less stable centers.\\

\begin{figure}[!htb]
    \centering
    \includegraphics[width=\linewidth]{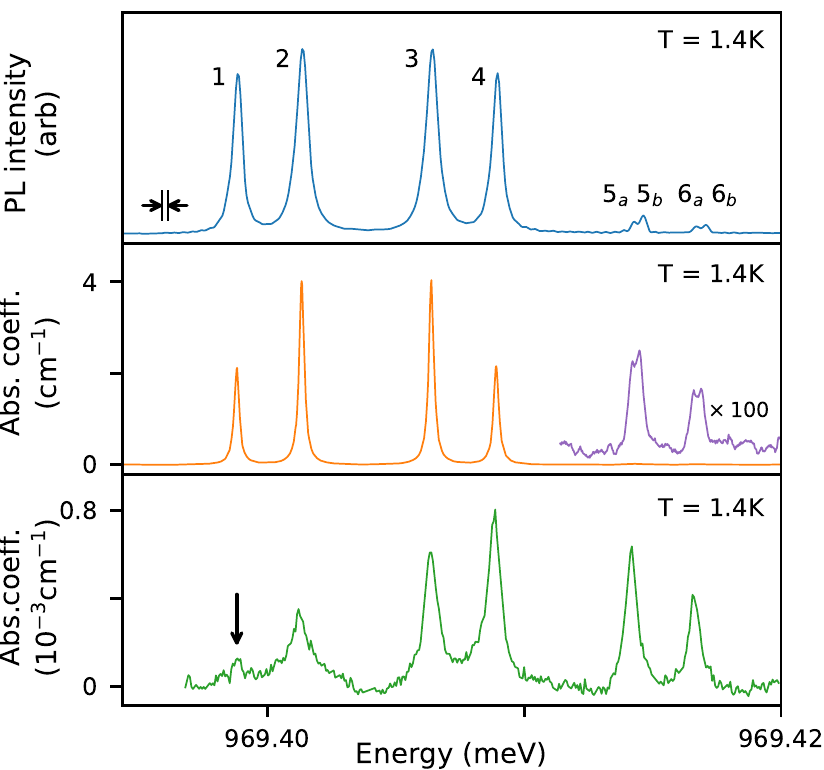}
    \caption{\textbf{(Top)} High resolution photoluminescence spectrum of the G-line in $^{28}$Si (\mbox{32 kGy}, annealed at \mbox{100 $^\circ$C}). The G-line splits into four main components, labeled 1 to 4. We attribute the smaller lines 5a, 5b, 6a, and 6b to centers where a $^{12}$C has been replaced by a naturally occuring $^{13}$C. The relative intensities of the G line components in luminescence are strongly modified by reabsorption. Spectrum recorded with a resolution of \mbox{0.31 $\upmu$eV} as indicated. \textbf{(Middle)} Absorption spectrum of the same transition for a $^{28}$Si sample with a \mbox{320 kGy} dose and \mbox{100 $^\circ$C} anneal, measured with a scanning single frequency laser. The $^{13}$C lines are enlarged for clarity. The sample was illuminated with low power above-gap radiation during the scan to prevent bleaching. \textbf{(Bottom)} Absorption spectrum on the same \mbox{320 kGy} $^{28}$Si sample with no above-bandgap illumination and a second higher power laser set to transition 1 (arrow) in order to bleach the $^{12}$C lines and reveal the full $^{13}$C quadruplet. The three spectra shown here were taken at a temperature of \mbox{1.4 K}.}
    \label{fig:G}
\end{figure}

High resolution spectra of the G center NP line taken in both PL and absorption are shown in \mbox{Fig. \ref{fig:G}}. The relative intensities of the components in the PL spectrum are strongly affected by reabsorption in the sample due to the remarkably strong G center absorption. Indeed, in the PL spectrum of the \mbox{320 kGy} sample (data not shown), the peaks labelled 1 through 4 are seen as dips in an apparently broadened spectrum. \\

The relative intensities of the G center components can be accurately determined from the absorption spectra, even for the \mbox{320 kGy} sample, where the absorption strengths were only approximately twice as large as in the \mbox{32 kGy} sample, perhaps indicating that the creation of G centers at \mbox{320 kGy} is limited by the availability of C$_\mathrm{s}$ in these $^{28}$Si samples. The absorption coefficients and integrated areas of the components are given in \mbox{Table \ref{tab:G}}. While G center absorption spectra were collected using the FTIR for all samples, there was a small contribution of the instrumental linewidth to the observed linewidth even at the highest available resolution. As a result, the absorption spectra shown in \mbox{Fig. \ref{fig:G}} were collected using a scanning DFB laser with a linewidth much less than that of the G-line components. The linewidths of the G center components are reported in \mbox{Table \ref{tab:PL}}; they are a little over twice the homogeneous linewidth associated with a \mbox{5.9 ns} lifetime \cite{Beaufils2018}, indicating that nearly all the inhomogeneous broadening has been removed in $^{28}$Si, and revealing the G center as a near-homogeneous light emitter in the bulk.\\

\begin{table}
  \centering
  \begin{tabularx}{\linewidth}{X X X}
  \hline \hline\noalign{\smallskip}
   & Abs. coeff. & Integrated area \\
   & [cm$^{-1}$] & [cm$^{-2}$] \\\noalign{\smallskip}\hline\noalign{\smallskip} %
   G$_1$ & 2.13 & 0.0060  \\ 
   G$_2$ & 4.02 & 0.0114  \\
   G$_3$ & 4.05 & 0.0117  \\ 
   G$_4$ & 2.17 & 0.0065  \\  
   G$_{5a}$ ($^{13}$C) & 0.015 & 0.00004  \\
   G$_{5b}$ ($^{13}$C) & 0.021 & 0.00007  \\  
   G$_{6a}$ ($^{13}$C) & 0.010 & 0.00003  \\ 
   G$_{6b}$ ($^{13}$C) & 0.013 & 0.00004  \\\noalign{\smallskip}\hline%\hline
  \end{tabularx}
    \caption{Peak absorption coefficients and integrated absorption areas for the G line components in the $^{28}$Si sample irradiated with \mbox{320 kGy} and annealed at 100$^{\circ}$C.}

  \label{tab:G}
\end{table}

The G-line in $^{28}$Si is seen to have a symmetrical quartet structure which has been completely hidden within the inhomogeneous linewidth seen before in $^{\mathrm{nat}}$Si. The relative intensities of the four components show within error a 1:2:2:1 ratio, and the pattern suggests transitions between two doublet levels with splittings of 2.5(2) $\upmu$eV and 7.5(2) $\upmu$eV, but we cannot at this time determine which of these splittings occurs in the ground state and which in the excited state, nor can we rule out transitions between singlet and quartet levels.\\

Also seen in \mbox{Fig. \ref{fig:G}} are weaker, higher-energy features labelled 5a, 5b, 6a and 6b, which we interpret as isotope-shifted versions of lines 3 and 4 for G centers which contain one $^{13}$C and one $^{12}$C instead of the predominant two $^{12}$C. The relative intensities of these components seen in the high resolution absorption spectrum are in reasonable agreement with the intensities to be expected from the $^{13}$C natural abundance \mbox{([ $^{13}$C]/[$^{12}$C] = 0.0111)}. If this identification of 5a through 6b is correct, there must be $^{13}$C versions of lines 1 and 2 hidden under lines 3 and 4, as will be verified later. The a-b doublet structure of lines 5 and 6 may result from the inequivalence of the two C atoms in the G center.\\

Isotope shifts of the unresolved NP G-line in $^\mathrm{nat}$Si due to one of the three isotopes of Si occupying the Si$_\mathrm{i}$ site have been previously reported \cite{Thonke1981,Davies1981} and are shown in the inset to the middle panel of \mbox{Fig. \ref{fig:all}}. The $^{29}$Si and $^{30}$Si peaks are obviously absent from the enriched $^{28}$Si spectrum. An unresolved C isotope shift of \mbox{40 $\upmu$eV} has been reported in comparing the relatively broad lines of $^\mathrm{nat}$Si samples containing primarily $^{12}$C or a mixture of $^{12}$C and $^{13}$C \cite{Davies1983}. The improved linewidths available in our $^{28}$Si spectra allow a much more precise measurement of the shifts due to one $^{13}$C, which for the a \{b\} component is \mbox{7.85(20) $\upmu$eV} \{\mbox{8.1(2) $\upmu$eV}\}.\\

When measuring the G-line absorption spectrum using the scanning DFB laser a gradual bleaching of the absorption intensity was noted. The rapidity of this bleaching scaled with the DFB laser intensity incident on the sample, and the bleaching could be reversed by illuminating the sample with above-gap light at 1047 nm. This may indicate that resonant excitation of a G-line component can drive the transition from the optically-active B state to the optically-inactive A state \cite{Song1990}, although the bleaching might also result from a change in charge state, or by populating a long-lived shelving state. The bleaching effect lasted at least 15 minutes with the sample not illuminated. On the ms to s time scales investigated here, pumping of either line 1, 2, 3 or 4 resulted in a simultaneous bleaching of all four main components.\\

Next we used this resonant bleaching effect to selectively reduce the intensity of the main $^{12}$C components to better reveal the spectrum of the isotope-shifted lines from centers having one $^{13}$C, as shown in the bottom panel of \mbox{Fig. \ref{fig:G}}. An additional higher power DFB laser was set to the energy of component 1, as indicated by the arrow, while the absorption spectrum was collected with a scanning lower power DFB laser. No above-gap excitation was used during this scan. This resonant bleaching greatly reduced the strength of lines 1 through 4, revealing the $^{13}$C-shifted versions of lines 1 and 2 which are almost coincident in energy with the $^{12}$C lines 3 and 4. Under these conditions all lines were somewhat broader, and the a-b splittings could no longer be resolved.\\

\begin{figure}[!htb]
    \centering
    \includegraphics[width=\linewidth]{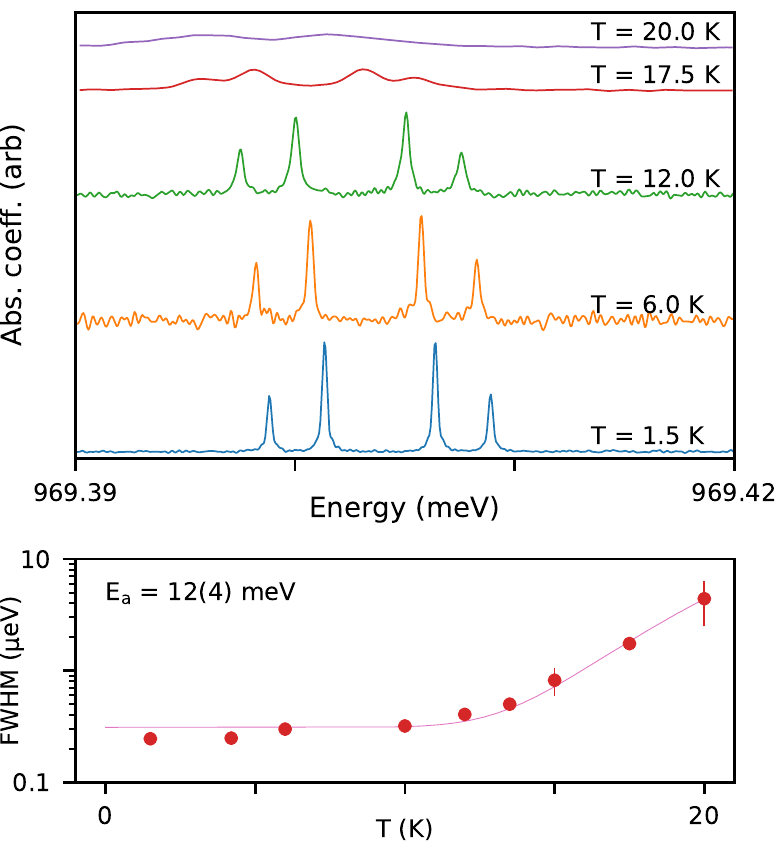}
    \caption{\textbf{(Top)} The G quadruplet components broaden identically with increasing temperature. \textbf{(Bottom)} The temperature dependence of the G linewidths is plotted versus temperature, from which is extracted an activation energy $E_a = 12(4)$meV. }
    \label{fig:G_T}
\end{figure}

The temperature dependence of the full width at half-maximum (FWHM) of the main G-line components is shown in \mbox{Fig. \ref{fig:G_T}}. The integrated absorption area of the combined G-line components was constant from 1.5 K to 20 K. Fitting the FWHM to a simple thermally-induced transitions model,

\begin{equation}
\mathrm{FWHM} = \mathrm{P}_0 + \frac{\mathrm{P}_\mathrm{T}}{\exp(\mathrm{E}_a/\mathrm{kT}) - 1}
\end{equation}

\noindent gave a thermal activation energy \mbox{$\mathrm{E}_\mathrm{a} = 12(4)$ meV} for this broadening, which may be evidence for an electronic excited state at that energy.\\

High resolution absorption spectra of the G-line components in $^{28}$Si were collected for applied magnetic fields up to \mbox{6 T}, and no changes from the zero field spectra were observed. Given the greatly reduced linewidths seen in $^{28}$Si, this confirms and strengthens earlier conclusions that the G-line arises from a spin zero to spin zero transition \cite{Thonke1981}.

\subsection{\label{subsec:C center}C center (790 meV, 1571 nm)}

The C center, with its NP line at \mbox{789.6 meV}, is one of the most thoroughly studied radiation damage centers in Si (for an earlier review see section 5.1 in \cite{Davies1989}, and for more recent reviews see \cite{Pajot2011} and \cite{Pajot2013}). It is known to have C$_\mathrm{1h}$ symmetry and to contain one C$_\mathrm{i}$ and one O$_\mathrm{i}$, and isotope shifts due to both C \cite{Davies1985, Kurner1989} and O \cite{Kurner1989} have been observed in spectra of the C center in $^\mathrm{nat}$Si. \\

\begin{figure}[!htb]
    \centering
    \includegraphics[width=\linewidth]{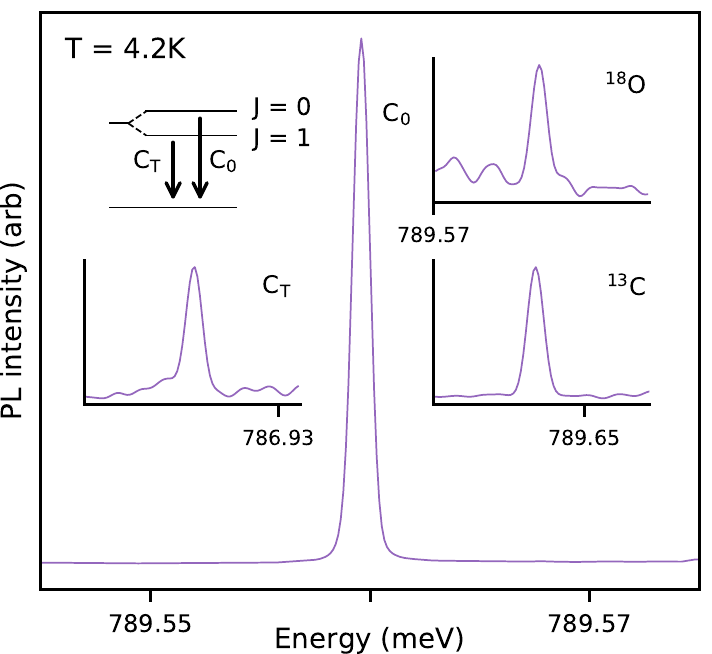}
    \caption{No-phonon C$_0$ photoluminescence in $^{28}$Si irradiated with \mbox{32 kGy} and annealed at \mbox{250$^\circ$C}. \textbf{(Top left inset)} Energy diagram showing the singlet and triplet bound exciton levels giving rise to the C$_0$ and C$_\mathrm{T}$ lines. \textbf{(Other insets)} The C$_\mathrm{T}$ line and the isotopically shifted C$_0$ lines resulting from naturally abundant $^{18}$O and $^{13}$C isotopes, with the same, but shifted, energy scale as the main figure. Relative to a normalized C$_0$-line, the amplitudes of the C$_\mathrm{T}$, $^{18}$O and $^{13}$C intensities are 0.0079, 0.0050 and 0.0267. Collected at \mbox{4.2 K} using \mbox{0.62 $\upmu$eV} resolution, with the linewidths dominated by the instrumental contribution.
}
    \label{fig:C}
\end{figure}

The electronic properties of the initial and final states of the C center are better understood than for the G and W centers. The C center is known to be a hole-attractive isoelectronic bound exciton (IBE), also known as a pseudodonor center, in which the hole is tightly bound by the defect potential, and an electron is Coulomb-bound to this positive charge \cite{Pajot2011}. The donor-like excited state spectrum of the C center has been observed using PLE spectroscopy \cite{Wagner1984}. In the pseudodonor IBE model, the tight binding of the hole is thought to quench its angular momentum. The resulting pseudospin \nicefrac{1}{2} hole coupled to the spin \nicefrac{1}{2} electron leads to an exchange splitting of the IBE ground state into singlet and triplet states, with the triplet state lying below the singlet. Since transitions from the IBE triplet state to the defect ground state are partially forbidden, the triplet IBE state has a much longer radiative lifetime than the singlet, and can be difficult to observe.\\

In a study of the transient decay of the C center PL, Bohnert et al. \cite{Bohnert1993} inferred that the observed NP C-line must originate from the singlet IBE state, and that there must be an unobserved triplet level lying \mbox{$\sim$3.2 meV} below the singlet level. This was confirmed later by the observation of PL from the triplet level (labelled C$_\mathrm{T}$) lying \mbox{2.64 meV} below the singlet transition (now labelled C$_0$) \cite{Ishikawa2009}. Here we find a singlet-triplet splitting of \mbox{2.633 meV}, in good agreement with the previously known result.\\

\begin{figure}[!htb]
    \centering
    \includegraphics[width=\linewidth]{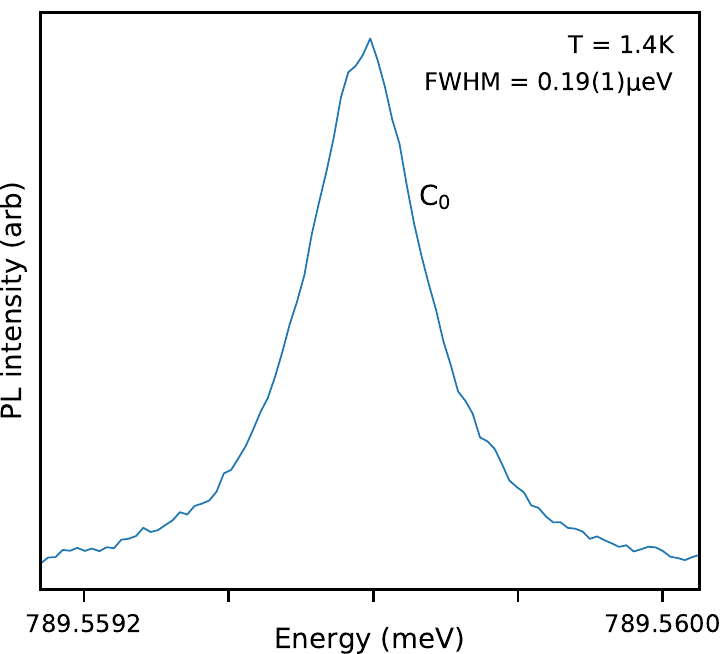}
    \caption{No-phonon C$_0$ photoluminescence in $^{28}$Si irradiated with \mbox{32 kGy} and annealed at \mbox{250$^\circ$C} measured in PLE with a scanning single frequency laser. 
}
    \label{fig:C_PLE}
\end{figure}

PL spectra collected with the FTIR of the main C$_0$-line together with the C$_\mathrm{T}$-line and $^{18}$O and $^{13}$C isotope-shifted versions of the C$_0$-line are shown in \mbox{Fig. \ref{fig:C}}. We know that the linewidth of the C$_0$-line was below the resolution limit of our FTIR, and there is no reason to believe that this was not also true of the C$_\mathrm{T}$ line and the isotope-shifted lines. The true linewidth of the C$_0$-line was determined by collecting a PLE spectrum using a scanning single frequency DFB diode laser, as shown in \mbox{Fig. \ref{fig:C_PLE}}, and reported in \mbox{Table \ref{tab:PL}}. Unfortunately, the SNR was insufficient to observe any of the other C components in PLE.\\

Two isotope-shifted C$_0$-lines are seen in \mbox{Fig. \ref{fig:C}}, one lying at \mbox{789.6477 meV} (also shown in \mbox{Fig. \ref{fig:all}}) that has been observed previously in natural silicon and attributed to the C center with its $^{12}$C replaced by a $^{13}$C \cite{Kurner1989}. We find an isotopic shift of \mbox{$\Delta$E = 0.0881 meV}, in good agreement with the previously published \mbox{$\Delta$E = 0.089 meV} value. We also observe a distinct isotopically shifted line due to $^{18}$O replacing $^{16}$O, with \mbox{$\Delta$E = 0.0154 meV}, which is reasonably close to the \mbox{0.024 meV} shift previously observed in $^\mathrm{nat}$Si samples with high $^{18}$O concentration \cite{Kurner1989}. The much narrower linewidths available in $^{28}$Si make possible a more accurate determination of these isotope shifts without requiring samples doped with isotopically-enriched O or C.\\

\subsection{\label{subsec:W center}W center (1018 meV, 1217 nm)}

The W NP line is known to be an electric dipole transition at a trigonal center having spin-zero initial and final states, and showing only a Si isotope shift in its local vibrational mode replicas \cite{Davies1987_2}. It is currently thought to consist of three Si$_\mathrm{i}$ \cite{Davies1987_2,Davies2006, Surma2008}, and shifted versions of the W-line have been observed in $^\mathrm{nat}$Si implanted with different noble gas ions \cite{Burger1984}. The W center is thought to exist near more heavily damaged regions, and its linewidth in $^\mathrm{nat}$Si is known to decrease with increasing annealing temperature \cite{Davies2006}.\\

\begin{figure}[!htb]
    \centering
    \includegraphics[width=\linewidth]{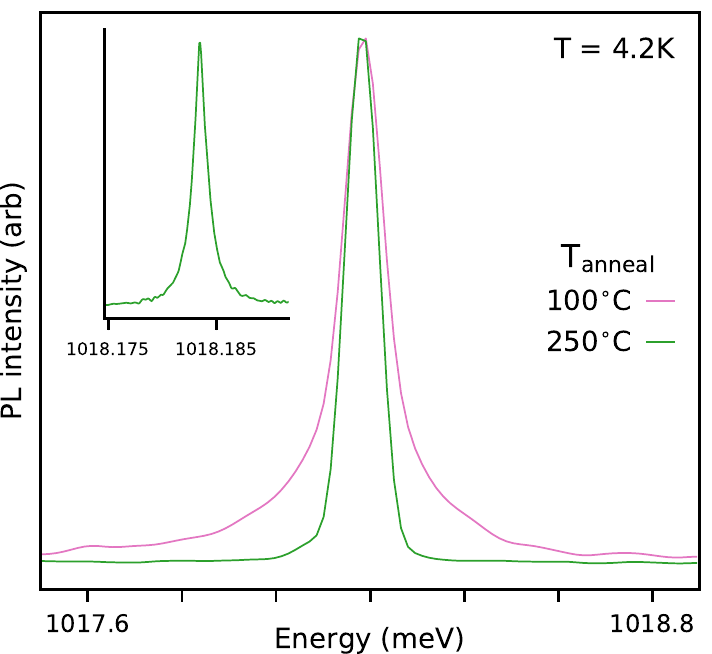}
    \caption{A low resolution (\mbox{62 $\upmu$eV)} comparison of the  normalized W-line photoluminescence in $^{28}$Si irradiated with \mbox{320 kGy} after anneals at \mbox{100 $^\circ$C} (pink) and \mbox{250 $^\circ$C} (green) shows the broader energy distribution of the W center for low anneal temperatures. (Inset) \mbox{0.62 $\upmu$eV} high resolution spectrum of the same line in a $^{28}$Si \mbox{32 kGy} \mbox{250 $^\circ$C} anneal sample.} 
    \label{fig:W}
\end{figure}

We observe this same behavior as shown in \mbox{Fig. \ref{fig:W}}, where the superimposed low resolution PL scans show a much broader energy distribution for the sample annealed at \mbox{100 $^\circ$C} compared to the sample annealed at \mbox{250 $^\circ$C}. Note that the linewidth of the \mbox{250 $^\circ$C} sample shown in the main panel of \mbox{Fig. \ref{fig:W}} is strictly instrumental; a high resolution PL scan of the sample annealed at \mbox{250 $^\circ$C} is shown in the inset, and it is from this scan that the linewidth reported in \mbox{Table \ref{tab:PL}} was obtained. Many weak, sharp lines were observed in the \mbox{100 $^\circ$C} annealed sample in the region immediately below the W-line, and while none of these agreed with known \cite{Burger1984} noble-gas-shifted W-lines, they may be W centers shifted by the presence of other nearby defects which are annealed out at \mbox{250 $^\circ$C}. In agreement with earlier observations, and models for the W center composition, no isotope shifted W NP lines were observed in $^{28}$Si.\\

\section{\label{sec:conclusion}Conclusion}

We have shown that the G, C and W radiation damage centers can readily be studied in highly-enriched, high-purity $^{28}$Si, even though the levels of C and O in the starting material are quite low. The inhomogeneous isotope broadening inherent to $^\mathrm{nat}$Si can be eliminated in $^{28}$Si, resulting in linewidth improvements of over two order of magnitude for the C and G centers. Indeed, the linewidth of the G center components seen in $^{28}$Si are only slightly more than twice the limit imposed by homogeneous lifetime broadening, given the measured \mbox{5.9 ns} G center lifetime \cite{Beaufils2018}. The improved linewidths in $^{28}$Si allow for the clear observation of distinct isotope-shifted components which are much more difficult to observe in $^\mathrm{nat}$Si. The spectrum of the G center in $^{28}$Si also reveals a quartet fine structure which is completely hidden in the inhomogeneously broadened spectrum in $^\mathrm{nat}$Si. This fine structure may have applications in quantum technology, and certainly has implications for models of the G center structure and initial and final states. These results also have direct implications for the use of individual centers as single photon emitters even in $^\mathrm{nat}$Si, since inhomogeneous isotope broadening will be absent for individual centers.\\

G, C and W are only a small subset of the known radiation damage centers in Si. The same linewidth improvements seen here for these centers in $^{28}$Si will apply to many other centers as well. The long list of known centers in silicon should be surveyed to look for possible applications in quantum technologies. In particular, it would be interesting to find a center which has an optically-resolved hyperfine splitting, since this could lead to an optically-accessible nuclear spin qubit, with the possibility of combining long coherence times with optical addressability, in a more accessible wavelength region than the proposed chalcogen spin/photon qubits which operate near \mbox{2900 nm} \cite{Morse2017}. The C center, with its convenient emission wavelength near 1550 nm and its potential transient access to a hyperfine-coupled nuclear spin via its excited state triplet, may provide a useful spin-photon interface. \\

\section*{Acknowledgements}
This work was supported by the Natural Sciences and Engineering Research Council of Canada (NSERC), the Canada Research Chairs program (CRC), the Canada Foundation for Innovation (CFI), and the B.C. Knowledge Development Fund (BCKDF). The $^{28}$Si samples used in this study were prepared from the Avo28 crystal produced by the International Avogadro Coordination (IAC) Project (2004-2011) in cooperation among the BIPM, the INRIM (Italy), the IRMM (EU), the NMIA (Australia), the NMIJ (Japan), the NPL (UK), and the PTB (Germany). We thank Alex English of Iotron Industries for assistance with the electron irradiations.


%\bibliography{mybib}
\end{document}